\documentclass{article}

\usepackage{arxiv}

\usepackage[utf8]{inputenc} % allow utf-8 input
\usepackage[T1]{fontenc}    % use 8-bit T1 fonts
\usepackage{hyperref}       % hyperlinks
\usepackage{url}            % simple URL typesetting
\usepackage{booktabs}       % professional-quality tables
\usepackage{amsfonts}       % blackboard math symbols
\usepackage{nicefrac}       % compact symbols for 1/2, etc.
\usepackage{microtype}      % microtypography
\usepackage{lipsum}		% Can be removed after putting your text content
\usepackage{graphicx}
\usepackage{natbib}
\usepackage{doi}
\usepackage{bbding}
\usepackage{array}
 \usepackage{array,multirow,graphicx}

\usepackage{amsmath}
\usepackage{diagbox}
\usepackage{color}
\usepackage{bm}
\usepackage{array}
\usepackage[utf8]{inputenc}
\usepackage{graphicx}
\usepackage{float}
\usepackage{amsmath}
\usepackage{caption}
\usepackage{subcaption}
\usepackage{multirow}
\renewcommand{\arraystretch}{2}
\usepackage{color}
\usepackage{booktabs}
\usepackage[ruled,vlined]{algorithm2e}
\usepackage{makecell}
\newcolumntype{I}{!{\vrule width 1.5pt}}
\usepackage{booktabs}

\usepackage{graphicx}  %needed to include png, eps figures
\usepackage{soul}
\usepackage{subcaption}
\usepackage{mwe}
\usepackage{xcolor}
\usepackage{amssymb}

\usepackage{tikz}
\definecolor{fc}{HTML}{1E90FF}
\tikzset{fc/.style={black,draw=black,fill=fc,rectangle,minimum height=1cm}}
\definecolor{h}{HTML}{228B22}
\definecolor{bias}{HTML}{87CEFA}
\tikzset{h/.style={black,draw=black,fill=h,rectangle,minimum height=1cm}}
\tikzset{bias/.style={black,draw=black,fill=bias,rectangle,minimum height=1cm}}
\definecolor{anti-flashwhite}{rgb}{0.95, 0.95, 0.96}
\definecolor{almond}{rgb}{0.98, 0.91, 0.71}

\title{A comprehensive comparison of neural operators \\ for 3D industry-scale engineering designs}
\author{
  Weiheng Zhong\textsuperscript{1}\thanks{Corresponding author: \href{mailto:weiheng4@illinois.edu}{weiheng4@illinois.edu}}
  \and
  Qibang Liu\textsuperscript{2}
  \and
  Diab Abueidda\textsuperscript{2}
  \and
  Seid Koric\textsuperscript{2,3}
  \and
  Hadi Meidani\textsuperscript{1}
}
\date{%
  \textsuperscript{1}\,Department of Civil and Environmental Engineering, University of Illinois Urbana-Champaign, Urbana, IL 61801\\
  \textsuperscript{2}\,National Center for Supercomputing Applications, University of Illinois Urbana-Champaign, Urbana, IL 61801\\
  \textsuperscript{3}\,Department of Mechanical Science and Engineering, University of Illinois Urbana-Champaign, Urbana, IL 61801
}

\begin{document}
\maketitle
\begin{abstract}
Neural operators have emerged as powerful tools for learning nonlinear mappings between function spaces, enabling real-time prediction of complex dynamics in diverse scientific and engineering applications. With their growing adoption in engineering design evaluation, a wide range of neural operator architectures have been proposed for various problem settings. However, model selection remains challenging due to the absence of fair and comprehensive comparisons. To address this, we propose and standardize six representative 3D industry-scale engineering design datasets spanning thermal analysis, linear elasticity, elasto-plasticity, time-dependent plastic problems, and computational fluid dynamics. All datasets include fully preprocessed inputs and outputs for model training, making them directly usable across diverse neural operator architectures. Using these datasets, we conduct a systematic comparison of four types of neural operator variants, including Branch-Trunk-based Neural Operators inspired by DeepONet, Graph-based Neural Operators inspired by Graph Neural Networks, Grid-based Neural Operators inspired by Fourier Neural Operators, and Point-based Neural Operators inspired by PointNet. We further introduce practical enhancements to adapt these models to different engineering settings, improving the fairness of the comparison. Our benchmarking study evaluates each model’s strengths and limitations in terms of predictive performance, computational efficiency, memory usage, and deployment complexity. The findings provide actionable insights to guide future neural operator development. All \textbf{source code} is available to the public at \href{https://github.com/WeihengZ/FC4NO/tree/main}{\textcolor{blue}{Github}}. The \textbf{data} is available at \textcolor{blue}{Harvard Dataverse} for the paper reviewing process: \href{https://dataverse.harvard.edu/dataset.xhtml?persistentId=doi:10.7910/DVN/OZQMD4}{Heat sink}, \href{https://dataverse.harvard.edu/dataset.xhtml?persistentId=doi:10.7910/DVN/3QKTEI}{Bracket}, \href{https://dataverse.harvard.edu/dataset.xhtml?persistentId=doi:10.7910/DVN/UWOYM2}{Bracket-time}, \href{https://dataverse.harvard.edu/dataset.xhtml?persistentId=doi:10.7910/DVN/XFUWJG}{JEB}, and \href{https://dataverse.harvard.edu/dataverse/DrivAerNet}{DrivAer and DriverNet++}
\end{abstract}

\keywords{Neural Operator, Fair Comparison, 3D Engineering Design, Industry-Scale Problems}

\section{Introduction \label{Sec.Intro}}

The numerical solution of partial differential equations (PDEs) is a foundational aspect of computational science and engineering, which can be used in instant evaluations in design, sensitivity analysis and uncertainty quantification, online controls, and digital twin development. Traditional discretization-based methods, including the finite element method (FEM) \citet{fem} and finite volume method (FVM) \citet{fvm}, have been extensively developed, offering high accuracy and robustness across various domains. However, despite their widespread use, these conventional approaches often incur substantial computational costs, especially for large-scale, high-fidelity simulations. In recent years, neural operators have emerged as a promising alternative, capable of learning mappings between infinite-dimensional function spaces directly from data~\cite{FNO,DON}. By bypassing traditional mesh discretization at inference time, neural operators significantly accelerate PDE solution prediction, unlocking new possibilities for real-time simulation, optimization, and control \citet{NO_review}.

One of the most critical applications of PDE solvers lies in engineering design evaluation, where rapid assessments of physical fields (e.g., temperature, stress, pressure) are essential for exploring a wide range of candidate designs \citet{NO_review}. In this context, neural operators offer the potential to significantly accelerate design iteration cycles. While many powerful neural operator architectures have been proposed, \textcolor{black}{existing evaluation frameworks have not been rigorously tested on industry-scale applications that demand both complex 3D geometries with high-resolution meshes (typically exceeding 10,000 nodes) and parametric representations—two fundamental characteristics of real-world engineering workflows. High-resolution 3D meshes are essential for accurately capturing the fine geometric details required in structural, thermal, and fluid simulations. Likewise, parametric representations are critical for manufacturability, as they enable design control through a small number of interpretable parameters such as length, thickness, or curvature \citet{parametricML, geom-DeepONet, parametricML2, parametricML3}.} However, most current benchmarks fall short of these requirements. First, the majority of datasets used to evaluate neural operators are predominantly 2D \citet{GNOT, wu2024transolver, scalingtransformers2023, D-FNO2025, NO_nonlocal, Neural_field, DANO, AROMA}, limiting their relevance for applications where full 3D modeling is essential. Second, existing evaluations often rely on free-form geometries \citet{li2023gino, wu2024transolver, GNOT}, which are rarely adopted in practical engineering due to manufacturing constraints. These mismatches between current evaluation practices and real-world design needs reveal a significant gap, underscoring the importance of developing more representative benchmarking frameworks for neural operators in industry-scale applications.

To bridge this gap, we introduce a standardized benchmarking framework for neural operators in 3D, parameterized, industry-scale engineering problems. Our approach begins with curating six diverse datasets that encompass a wide range of engineering challenges, including thermal analysis, linear elasticity, elasto-plastic deformation, time-dependent plasticity, and computational fluid dynamics. These datasets are designed to capture the complexity of real-world applications. We then systematically classify state-of-the-art neural operator architectures into four major categories—Branch-Trunk-based \citet{DON, S-DON, geom-DeepONet, DCON, GANO}, Graph-based \citet{GNO, EAGNO}, Grid-based \citet{li2023gino, figconvunet}, and Point-based operators \citet{GNOT, pointnet, wu2024transolver}. From each category, we identify representative models and ensure their fair adaptation and evaluation on all datasets. Our benchmarking process comprehensively assesses these models, focusing on predictive accuracy, computational efficiency, memory usage, and deployment practicality. Our contributions are summarized as follows:
\begin{itemize} 
\item We curated six industry-level 3D engineering datasets, enabling a wide range of neural operators training and evaluation. 
\item We present a comprehensive taxonomy of neural operator architectures, adapting each model to both parametric and free-form geometries and proposing architectural enhancements to ensure fair comparisons.
\item We conduct the first extensive benchmarking study of neural operators on diverse industry-scale engineering designs, providing actionable insights for future neural operator development and deployment.
\item We release all datasets, model implementations, and evaluation code, promoting reproducibility and facilitating further research in neural operators for industry-scale engineering design. 
\end{itemize}

 \section{Related works}

\subsection{DeepONet-inspired Neural Operator}

Deep Operator Networks (DeepONets), introduced by \cite{DON}., are neural architectures designed to learn nonlinear operators—mappings between infinite-dimensional function spaces—enabling efficient approximation of complex physical models from limited data \cite{Lu2021,Lu2019}. Their core structure consists of a branch network encoding input functions at fixed sensor locations and a trunk network processing output coordinates, with outputs combined via a dot product \cite{Lu2021}. Supported by a universal operator approximation theorem, DeepONet generalizes well to nonlinear and parametric PDEs, while physics-informed variants incorporate governing equations directly into the loss to avoid numerical solvers \cite{Lu2021,Yang2024}.

To address scalability and uncertainty, Separable DeepONet uses factorized subnetworks to mitigate the curse of dimensionality \cite{Mandl2024}, and $\alpha$-VI DeepONet adopts variational inference for robust uncertainty quantification \cite{Wang2024}. Computational efficiency is further improved via RaNN-DeepONet’s randomized networks \cite{Zhang2025}, derivative-enhanced losses \cite{Zhang2024}, and Point-DeepONet’s mesh-free learning on 3D point clouds \cite{Park2025}.

Applications span porous media flow, carbon sequestration, and PDE-constrained control. Architectures like U-DeepONet employ U-Net backbones to improve multiphase flow modeling \cite{Diab2024,Li2024}, while nonlinear reconstructions extend DeepONet to PDEs with discontinuities \cite{Lanthaler2022}. Transfer learning techniques enhance adaptability across problem settings, enabling efficient fine-tuning for multi-operator tasks \cite{Goswami2022,Yadav2023}. Overall, DeepONet remains a powerful and evolving tool for physics-informed machine learning.

\subsection{Graph-based Neural Operator}

Graph Neural Operators (GNOs) have emerged as powerful tools for learning mappings between function spaces defined on irregular domains, particularly for solving partial differential equations (PDEs). A pivotal example is the Multipole Graph Neural Operator (MGNO) \cite{li2020multipole}, which incorporates a hierarchical, multi-level graph representation to capture long-range interactions in PDE solutions efficiently. This approach reduces computational complexity from quadratic to linear in the number of nodes, thereby enabling scalable and discretization-invariant GNO architectures. Building upon the GNO framework, the Latent Neural Operator (LNO) \cite{lno2024} encodes inputs into a latent graph space enriched by a Physics-Cross-Attention mechanism to learn more expressive graph-based operator mappings. CORAL \cite{serrano2023coral} extends GNOs by using coordinate-based neural representations, allowing operators to generalize to PDEs defined on arbitrary and non-mesh-conforming geometries. The Geometry-Informed Neural Operator (GINO) \cite{li2023gino} further integrates graph-based and spectral operator learning to construct latent regular grids from unstructured 3D domains for solving large-scale PDEs. PDE-GCN \cite{eliasof2021pdegcn} exemplifies a graph neural operator architecture derived from PDE discretizations, which combats over-smoothing in deep GNN layers through PDE-aware message passing. To improve numerical stability, the Representation Equivalent Neural Operator (RENO) framework \cite{serrano2023repneuralops} introduces alias-free GNOs by aligning discrete GNN computations with the properties of continuous operators. Scalable Transformer variants adapted for graph operator learning \cite{scalingtransformers2023} allow GNOs to handle high-dimensional PDE surrogate modeling tasks while maintaining strong accuracy and efficiency. Graph-Tuple Neural Networks (GtNNs) \cite{velasco2024gtnn} expand GNOs to multimodal node and edge features with theoretical guarantees for stability under perturbations. Lastly, the Alias-Free Mamba Neural Operator (MambaNO) \cite{zheng2024mambano} achieves linear-complexity GNO inference by combining long-range dependency modeling and implicit time integration, offering a new paradigm for graph-based operator learning.

\subsection{Grid-based Neural Operator}

Fourier Neural Operators (FNOs) represent a groundbreaking approach in neural operator learning, efficiently approximating solution operators of parametric partial differential equations (PDEs) by operating in the frequency domain \cite{Li2020}. They utilize Fourier transforms with learned spectral filters to capture global dependencies and achieve resolution-invariant learning with quasi-linear time complexity, excelling in modeling complex systems such as turbulent flows and fluid dynamics \cite{Li2020,Li2021}. The architecture lifts input functions to higher-dimensional representations, applies Fourier domain operations, and reconstructs target functions, providing high accuracy and computational efficiency compared to classical solvers \cite{Li2020}.

Extensions enhancing FNO scalability, flexibility, and applicability include multi-grid architectures like MgFNO and Multi-Grid Tensorized FNO (MG-TFNO), which leverage hierarchical frequency decompositions to improve generalization and reduce memory usage for large-scale PDE problems \cite{MGTFNO2024,D-FNO2025}. Domain Agnostic FNOs (DAFNO) address irregular grids and diverse domain topologies, while learnable Fourier filters enable more localized spectral feature extraction \cite{DAFNO2023,GaborFNO2024}. Hybrid models integrating FNOs with CNNs or graph neural operators combine local spatial and global spectral information for enhanced predictive performance \cite{LocalizedKernel2024}. Variants such as U-FNO and HyperFNO demonstrate superior speed, accuracy, and data efficiency in multiphase flow and PDE simulations \cite{U-FNO2022}.

Theoretical analyses confirm FNOs' universal approximation capabilities for continuous operators and convergence under discretization refinements \cite{UniversalFNO2021}. Physics-informed training regimes incorporate PDE constraints directly into learning to improve accuracy and stability \cite{PINO2021}. Extensive benchmarks demonstrate FNOs outperform classical numerical solvers with speed-ups exceeding 100×, while retaining high fidelity in fluid dynamics, Darcy flows, and turbulent simulations \cite{BenchmarksFNO2021}. FNOs and their innovations thus form a versatile and efficient framework pivotal to advancing scientific machine learning.

\subsection{Point-based Neural Operator}

Point-based neural operator models have emerged as a powerful approach for learning solution operators of partial differential equations (PDEs) by directly operating on spatial point clouds. This representation enables flexibility in handling irregular geometries, complex boundary conditions, and variable discretization. The Geometry-Informed Neural Operator (GINO) introduced by Li et al.~\cite{li2023geometry} exemplifies this approach by combining graph neural networks on point clouds with Fourier Neural Operators acting on latent regular grids derived from geometric encodings such as signed distance functions. GINO achieves remarkable computational acceleration and discretization invariance, enabling efficient and accurate surrogate modeling for large-scale 3D PDEs in industrial scenarios.

Transformer-based architectures further advance point-based modeling of PDEs. ~\cite{li2023scalable} and ~\cite{wu2024transolver} propose scalable pointwise transformers—such as Transolver—that utilize attention mechanisms over spatial points, effectively learning solution operators on general geometries with arbitrary discretizations. The Universal Physics Transformers (UPTs) framework~\cite{alkin2024universal} unifies spatio-temporal PDE modeling with shared transformer architectures acting on point-level data, demonstrating versatility and computational efficiency~\cite{alkin2024universal, yu2024nonlocal}. Complementarily, coordinate-based neural field methods like CORAL~\cite{serrano2023operator} bypass traditional meshing by learning operator mappings directly from point samples, preserving spatial structure and handling complex geometries robustly.

Recent innovations include alias-free and nonlocal operator models that use pointwise attention to uncover interpretable physics and mitigate numerical artifacts, as seen in the Alias-Free Mamba Neural Operator~\cite{zheng2024alias} and the Nonlocal Attention Operator~\cite{yu2024nonlocal}. Graph-based models such as GFN~\cite{morrison2024gfn} and inducing point transformer methods~\cite{lee2024inducing} also operate on point discretizations to ensure resolution invariance and scalability. Enhanced operator learning incorporates derivative information~\cite{qiu2024derivative} and Bayesian uncertainty quantification, improving accuracy and robustness. Collectively, these point-based neural operators provide flexible, scalable, and accurate methods for PDE surrogate modeling on complex domains, pushing forward the frontiers of computational scientific modeling.

\section{Evaluation framework on industry-scale engineering problems \label{Sec.data}}

A neural operator learns a mapping between infinite-dimensional function spaces. Let $ D \subset \mathbb{R}^d $ be a Lipschitz domain \citet{lipschitz}, and let $ \mathcal{B} $ be a Banach space of real-valued functions defined on $ D $. Consider a family of level set functions \citet{level_set_fun} $\mathcal{S} \subset \mathcal{B} $, where each $ s \in \mathcal{S} $ defines a $d$-dimensional submanifold $ \Omega_S = \{ x \in D : s(x) > 0 \} $. Given a partial differential operator $ \mathcal{L} $, we study the solution $ u \in \mathcal{U}_T \subset \mathcal{B} $ of the following PDE:
\begin{align}
    \mathcal{L}(u(x)) &= f(x), \quad x \in \Omega_S, \\
    u(x) &= g(x), \quad x \in \partial \Omega_S,
\end{align}
where $ f \in \mathcal{F} \subset \mathcal{B} $ denotes the PDE forcing term, $ g \in \mathcal{G} \subset \mathcal{B} $ represents boundary and initial conditions, and $ x \in D $ is the spatial coordinate. The neural operator seeks to approximate the solution operator as a mapping from PDE parameters and domain geometries to PDE solutions:
\begin{equation}
    \mathcal{O}: \mathcal{F} \times \mathcal{G} \times \mathcal{S} \rightarrow \mathcal{U},
\end{equation}
where $ \mathcal{U} $ is a Banach space representing the family of PDE solutions on varying domains $ \Omega_S $. In practice, neural operator models are trained on a finite collection of discrete observations of these functions~\cite{FNO}. In many engineering optimization problems, parametric representations of PDE parameters (e.g., loading, inlet velocity) and design geometries are available. Accordingly, we predict the PDE solution with a neural operator model $ M $ in the following setting:
\begin{equation}
    u(x_q) = M(x_q, F, G, S, p),
\end{equation}
where $x_q \in D$ is a query point at which the PDE solution is predicted, $ F = \{(x_i, f(x_i))\}_{i=1}^m $, $ G = \{(x_i, g(x_i))\}_{i=1}^n $, and $ S = \{x_i\}_{i=1}^l $ represent the point cloud representations of the parameter $ f $, the boundary/initial conditions $ g $, and the geometry $ \Omega_S $, respectively, and the vector $ p \in \mathbb{R}^q $ denotes a $ q $-dimensional parametric representation of the design space. We evaluate model performance across six real-world engineering datasets, covering a diverse range of physics and geometry representations:

\begin{itemize} 
\item \textbf{Heatsink:} A thermal conduction dataset simulating heat dissipation from a CPU via a heat sink. The geometry is parameterized by fin height, fin thickness, fin spacing, and the number of fins. Boundary conditions include force convection on the side surfaces and free convection on the top. The steady-state temperature field is computed as the PDE solution.
\item \textbf{Bracket:} This dataset models an elasto-plastic deformation problem in mechanical brackets with parameterized geometry, including length, thickness, and hole radius. Each sample also varies in applied pressure loading. FEM simulations use quadratic hexahedral elements, an elastic-plastic constitutive law with isotropic hardening, and small deformation assumptions.
\item \textbf{Bracket-time:} A time-dependent extension of the bracket dataset, where varying loading sequences are applied over 101 time steps. The goal is to learn the final stress state of the bracket in elasto-plastic modeling.
\item \textbf{JEB:} The Jet Engine Bracket (JEB) dataset consists of 2,138 geometrically diverse 3D bracket designs sourced from the GE Jet Engine Bracket Challenge \citet{deepJEB}. The shapes exhibit significant freeform variation without a shared parameterization. The task is to predict the von Mises stress field under a prescribed vertical load.
\item \textbf{DrivAer:} A large-scale CFD dataset offering high-fidelity aerodynamic simulations across multiple 3D vehicle geometries with parametric variation \citet{drivAer}. The objective is to predict surface pressure distribution under standardized wind flow conditions.
\item \textbf{DrivAer++:} An extension of DrivAerNet involving freeform, non-parametric vehicle geometries \citet{drivAer++}. This dataset focuses on evaluating model generalization to unparameterized, complex CAD geometries in automotive fluid dynamics.
\end{itemize}

\begin{figure}[h]
    \centering
    \begin{subfigure}[b]{0.45\textwidth}
        \includegraphics[width=\textwidth]{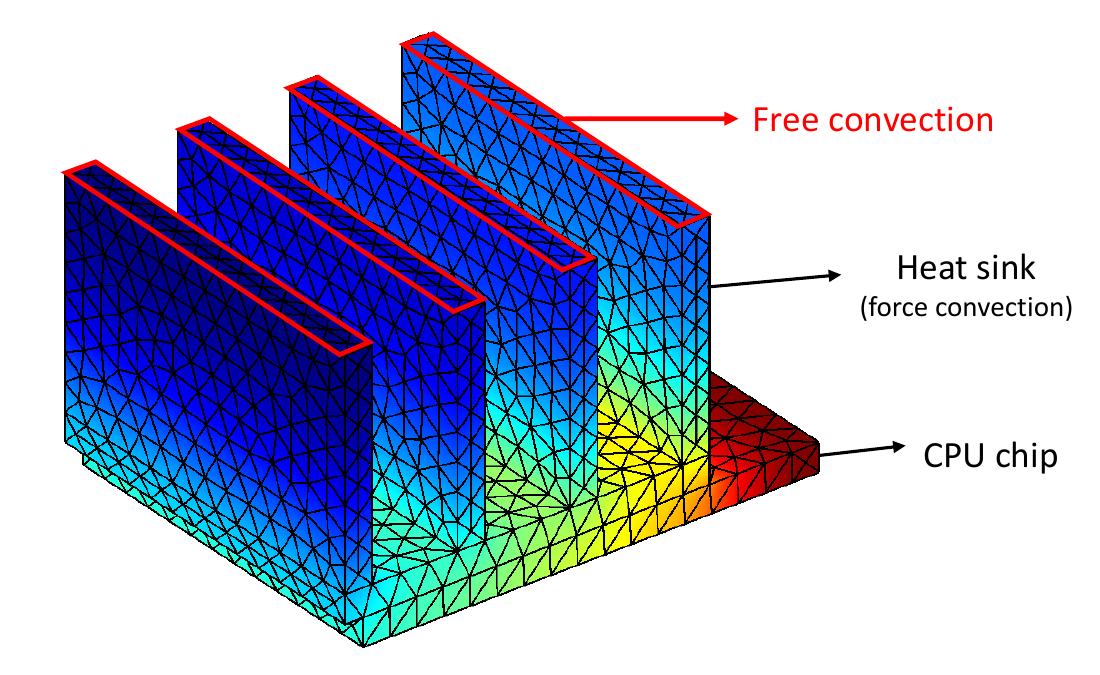}
        \caption{Heat sink simulation}
    \end{subfigure}
    \begin{subfigure}[b]{0.35\textwidth}
        \includegraphics[width=\textwidth]{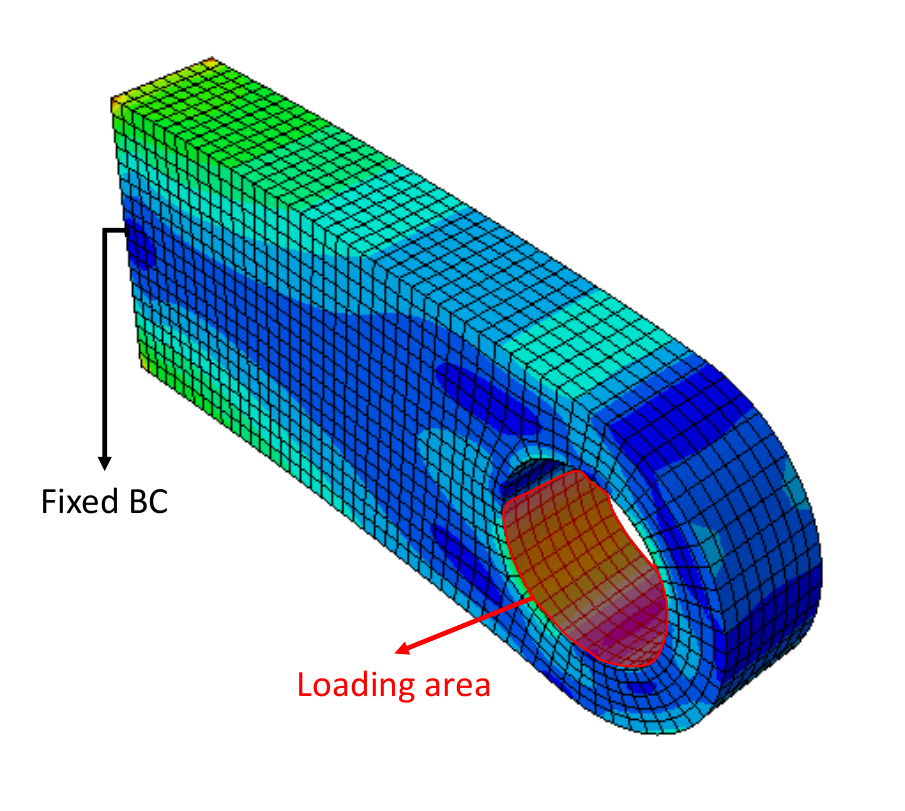}
        \caption{Bracket simulation}
    \end{subfigure}
    \begin{subfigure}[b]{0.4\textwidth}
        \includegraphics[width=\textwidth]{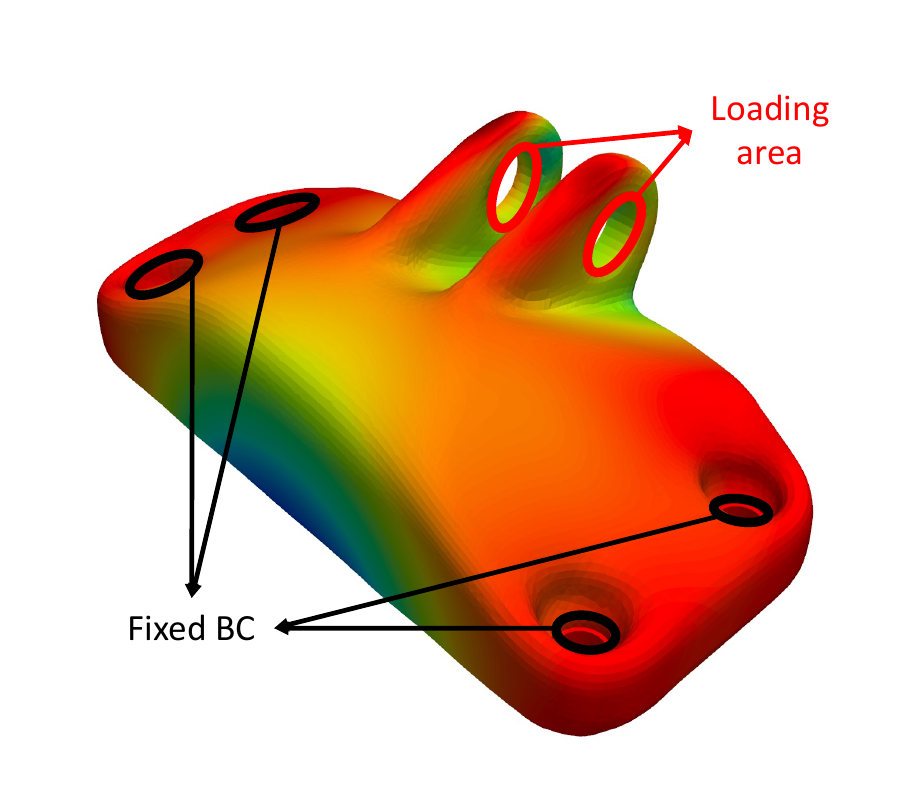}
        \caption{JEB dataset}
    \end{subfigure}
    \begin{subfigure}[b]{0.4\textwidth}
        \includegraphics[width=\textwidth]{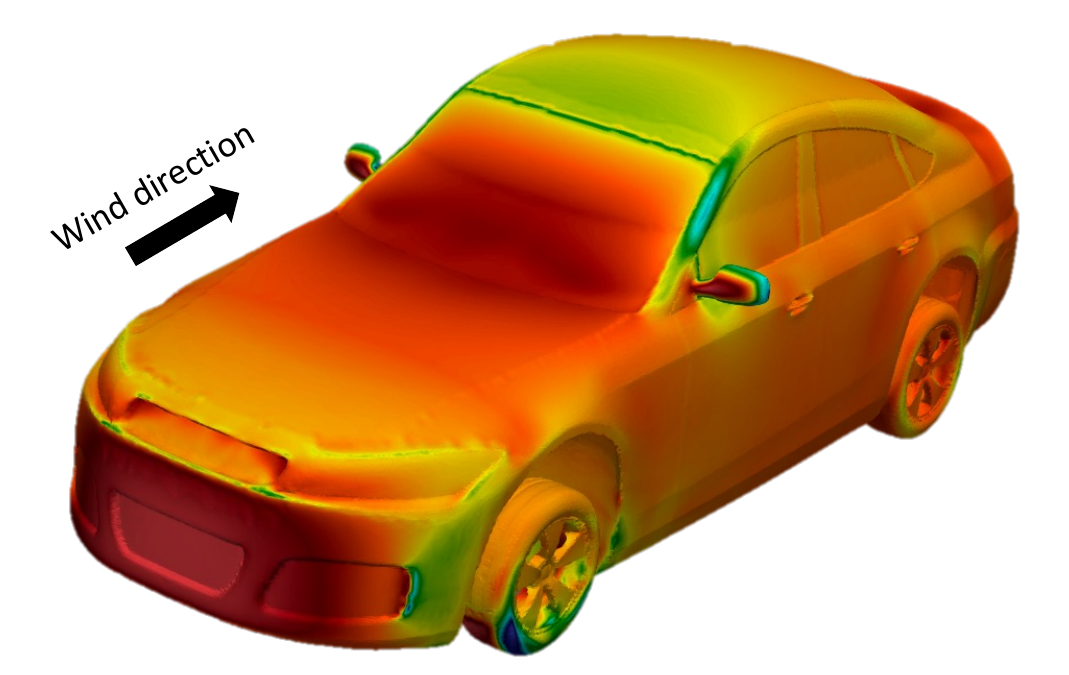}
        \caption{CFD simulation}
    \end{subfigure}
    \caption{\footnotesize Branch-trunk model architecture enhancements for freeform geometry.}
    \label{fig.data_visual}
\end{figure}

In summary, each dataset is a representation of one type of engineering problem, which is summarized in Table \ref{tab:dataset_summary} The Visualization of the sample geometry and PDE solutions are shown in \ref{fig.data_visual}.

\begin{table}[h]
\centering
\caption{Summary of datasets with graph size and associated challenges.}
\label{tab:dataset_summary}
\begin{tabular}{|c|c|c|c|c|c|}
\hline
\textbf{Name} & \textbf{Type} & \textbf{Data Size} & \textbf{Mesh Size}  & \textbf{\makecell{Geometry \\ Representation}} & \textbf{Challenges} \\
\hline
Heat sink & Thermal & 625 & $\sim$10k & Parametric &  Varying design geometry\\
\hline
Bracket & \makecell{Plastic \\ mechanics} & 3,000 & $\sim$10k & Parametric & Varying geometry and loading \\
\hline
Bracket-dynamic & \makecell{Plastic \\ mechanics}  & 10,000 & $\sim$30k & N/A (Fixed geometry) & Time-dependent loading \\
\hline
JEB & \makecell{Elastic \\ mechanics}  & 2,000 & $\sim$100k  & Non-parametric & \makecell{Small set of \\ complex geometry samples} \\
\hline
DrivAerNet & CFD & 4,000 & $\sim$500k  & Parametric & Large mesh \\
\hline
DrivAerNet++ & CFD & 4,000 & $\sim$400k & Non-parametric & \makecell{Large mesh with \\ non-parametric geometry}\\
\hline
\end{tabular}
\end{table}

\section{Neural operator architectures \label{Sec.NO}}

\subsection{Branch-Trunk-Based Neural Operators}

As one of the earliest neural operator architectures, Deep Operator Networks (DeepONet)~\cite{DON} leverage the universal approximation theorem for operators by representing the PDE solution as an inner product between two neural networks: the branch network and the trunk network. The predicted solution at a query location $x_q $ is given by:
\begin{equation}
    u(x_q) = \text{Sum}(b \odot t) = \text{Sum}(\mathcal{U}_b(p) \odot \mathcal{U}_t(x_q)),
\end{equation}
where $\mathcal{U}_b $ is a multi-layer perceptron (MLP) applied to the input parameter $p $, $\mathcal{U}_t $ is an MLP applied to the spatial query $x_q $, and $\odot$ represents the element-wise multiplications. However, the simplicity of this architecture limits its effectiveness in addressing complex engineering problems. For model performance improvement, Geom-DeepONet \citet{geom-DeepONet} replaces the branch MLP with a sinusoidal representation network (SIREN). On the other side, the Deep Compositional Operator Network (DCON) \citet{DCON} increases model expressiveness by introducing a hierarchical composition of operator layers, For example, with two layers, the solution is approximated by:
\begin{equation}
    u(x_q) = \mathcal{U}_u^2 \left( b \odot \mathcal{U}_u^1(b \odot t) \right),
\end{equation}
where $\mathcal{U}_u^1 $ and $\mathcal{U}_u^2 $ are learnable MLPs. As for problems with time-dependent PDEs, S-DeepONet \citet{S-DON} replaces the branch MLP with a gated recurrent unit (GRU), enabling temporal sequence encoding in the branch net:
\begin{equation}
    b = \text{GRU}([f(t_0), \ldots, f(t_m)]),
\end{equation}
where $f(t_k) $ denotes a time-dependent scalar function values at time step $t_k$. To enable handling of free-form geometries, Geometry-aware Neural Operator (GANO) \citet{GANO} extends DCON by incorporating point-cloud-based geometric encoding, where a global feature vector $E_g $ is concatenated with the trunk input:
\begin{align}
    u(x_q) &= \mathcal{U}_u^2 \left( b \odot \mathcal{U}_u^1(b \odot [t \parallel E_g]) \right), \\
    E_g &= \text{Pooling}(\{ \mathcal{U}_g(x_i) \}_{i=1}^l),
\end{align}
where $\parallel $ denotes concatenation, and $\mathcal{U}_g $ encodes each geometry point $x_i $. Overall, compared with traditional DeepONet, these improved branch-trunk models provide complementary strengths for different engineering problems. 

\subsection{Graph-Based Neural Operators}

Graph Neural Operators (GNOs) \citet{GNO} approximate mappings between function spaces using message passing over graphs, making them suitable for solving PDEs on irregular meshes. A typical GNO predicts the PDE solution on the mesh point coordinate $x_i$ with stacked graph convolutions:
\begin{equation}
    u(x_i) = \text{GC}^l \circ \text{GC}^{l-1} ... \circ \text{GC}^{1} (v_i), 
\end{equation}
where $v$ is the node features in a graph, and each graph convolution layer $\text{GC}$ is defined by: 
\begin{equation}
    v^{l+1}_i = \text{GC}(v^l_i) = \sum_{j \in \mathcal{N}_v} \mathcal{U}_{\text{mp}}(v^l_j, v^l_i)\, \mathcal{U}_f(v^l_j),
\end{equation}
where $\mathcal{N}_v$ is the set of neighbor nodes, $\mathcal{U}_{\text{mp}} $ is the message passing function, and $\mathcal{U}_f $ encodes the neighbor features. Edge-Augmented GNNs (EA-GNNs) \citet{EAGNO} improve long-range information flow by introducing edges between distant node pairs. These are often heuristically or randomly added in point clouds, reducing graph diameter. From a deployment perspective, both models are practical for engineering applications, as they can directly leverage existing meshes generated from finite element analysis without requiring additional pre-processing.

\subsection{Grid-Based Neural Operators}

Fourier Neural Operator (FNO) \citet{FNO} initiates the grid-based operator learning approach by performing convolution in the spectral domain, which predicts the PDE solution on a grid by:
\begin{equation}
    U = \mathcal{U}^l_{\text{FL}} \circ \mathcal{U}^{l-1}_{\text{FL}} ... \circ \mathcal{U}^1_{\text{FL}} (X),
\end{equation}
where $U$ is the PDE solution function values evaluated on a set of uniform grid coordinates $X$, and the Fourier Layer $\mathcal{U}_{\text{FL}}$ is formulated by:
\begin{equation}
    G^{l+1} = \mathcal{U}_{\text{FL}}(G^l) = \mathcal{F}^{-1}(R_{\phi} \cdot \mathcal{F}(G^l)),
\end{equation}
where $G^l$ is the grid feature output of the $l$-th Fourier layer, $\mathcal{F} $ and $\mathcal{F}^{-1} $ denote the Fourier and inverse Fourier transforms, respectively, and $R_\phi $ is a learned spectral filter. Geometry-Informed Neural Operator (GINO) \citet{li2023gino} extends FNO by integrating point cloud information directly. It uses a Riemann-sum-based discretization to construct the mapping between unstructured point features and grid-aligned features:
\begin{align}
    v^1_g &= \sum_{\|x_g - x_i\|_2 \leq r} \mathcal{U}_{\text{mp}}(v^0_i, v^0_g)\, \mathcal{U}_f(v^0_i), \quad v^1_g \in G^1, \quad v^0_g \in G^0, \quad v^0_i \in X, \\
    G^l &= \mathcal{U}^l_{\text{FL}} \circ \dots \circ \mathcal{U}^1_{\text{FL}}(G^1), \\
    v^l_i &= \sum_{\|x_g - x_i\|_2 \leq r} \mathcal{U}_{\text{mp}}(v^l_g, v^0_i)\, \mathcal{U}_f(v^l_g), \quad v^l_i \in X^l, \quad v^l_g \in G^l,
\end{align}
where $r$ is the hyper-parameter to search the neighbor nodes of each grid point, $v_i $ and $v_g $ represent node and grid features, respectively. However, storing and processing a high-resolution 3D grid is computationally expensive and memory inefficient. To alleviate this, FigConvNet \citet{figconvunet} uses factorized implicit grids instead of explicit 3D grids and processes data with a U-Net backbone, which drastically improves memory efficiency and has demonstrated strong performance on large CFD meshes.  From a deployment perspective, although the method requires radius search operations, these can be efficiently executed using hash grid algorithms \citet{hashgrid} available in GPU-accelerated libraries such as Warp. Therefore, we consider that both GINO and FigConvNet are suitable for large-scale engineering applications.

\subsection{Point-Based Neural Operators}

Point-based neural operators aim to directly model solution operators on point clouds. They offer high flexibility and are naturally suited to domains with nonuniform resolution and geometry. PointNet \citet{pointnet} processes unordered sets using shared MLPs and symmetric pooling, which is easily deployed but lacks local structure modeling. More recently, transformer-based architectures \citet{GNOT} have emerged, predicting PDE solution for one point location with stacked attention layers:
\begin{equation}
    u(x_i) = \text{Attn}^l \circ \text{Attn}^{l-1} ... \circ \text{Attn}^{1} (v_i),
\end{equation}
where the attention mechanism is formulated as:
\begin{align}
    v^{l+1}_i &= \text{Attn}(v_i^l) = \sum_j \frac{\exp(Q^l_i \cdot K^l_j / \tau)}{S_i} \cdot V^l_j, \\
    S_i &= \sum_j \exp(Q^l_i \cdot K^l_j / \tau),
\end{align}
where $\tau$ is the hyper-parameter, $Q^l_i = \mathcal{U}_q(v^l_i) $, $K^l_j = \mathcal{U}_k(v^l_j) $, and $V^l_j = \mathcal{U}_v(v^l_j) $ represents the queries, keys, and values of the point features. To further enhance the efficiency of transformer nerual operators, Transolver \citet{wu2024transolver} proposes a linear self-attention mechanism using physics-aware slice tokens. Each point feature $v_j $ is  assigned to $M $ slices using a MLP $\mathcal{U}_{\text{slice}}$:
\begin{align}
    \{w^i_j\}_{i=1}^M &= \{\text{Softmax}\left( \mathcal{U}_{\text{slice}}^i(v_j) \right)\}_{i=1}^M, \\
    s^i &= \frac{\sum_j w^i_j v_j}{\sum_j w^i_j},
\end{align}
where $s^i$ is the $i$-th physics token. After the stacked linear attention updates of the physics tokens, the point features is computed by composing the weights and the physics tokens:
\begin{align}
    s^{i,l} &= \text{Attn}^l \circ \dots \circ \text{Attn}^1(s^{i,0}), \\
    u(x_j) &= \sum_i w^i_j \cdot s^{i,l}.
\end{align}
Among existing point-based neural operators, we consider that PointNet, GNOT, and Transolver are particularly well-suited for engineering applications, as they can operate directly on raw point cloud data without the need for additional pre-processing.

\section{Architectural Enhancement Strategies}

Based on the categorization in Section~\ref{Sec.NO}, we refer to graph-based, grid-based, and point-cloud-based neural operators as geometric neural operators, as each operates on specific geometric representations. In contrast, branch-trunk architectures are generally designed to model global function spaces, relying heavily on parametric representations of the domain. To ensure a fair and consistent comparison among all selected neural operator models, we need to address several architectural challenges:

\begin{itemize}
    \item Many branch-trunk models are limited to problems with explicit parametric representations. Their ability to generalize to freeform geometries remains an open question.
    \item For problems with parametric design representations, geometric neural operators may be at a disadvantage due to the absence of explicit encoding of these parameters. This makes direct comparison to branch-trunk models potentially unfair.
\end{itemize}

\subsection{Branch-Trunk-Based Models for Freeform Geometries}

To extend branch-trunk-based models to support freeform geometries, we introduce two strategies for geometry encoding within the branch network:

\begin{enumerate}
    \item \textbf{High-level geometric descriptors:} We compute statistical features from the point cloud representation of the geometry, including centroid, bounding box dimensions, and PCA-derived axes and eigenvalues. These descriptors form a compact feature vector representing the global geometry.
    
    \item \textbf{Voxel-based encoding:} The geometry is discretized into a voxel grid, which is processed by a 3D convolutional neural network (CNN) to extract a latent embedding. This representation captures the spatial structure of the geometry in a resolution-aware manner.
\end{enumerate}
\begin{figure}[h]
    \centering
    \begin{subfigure}[b]{0.45\textwidth}
        \includegraphics[width=\textwidth]{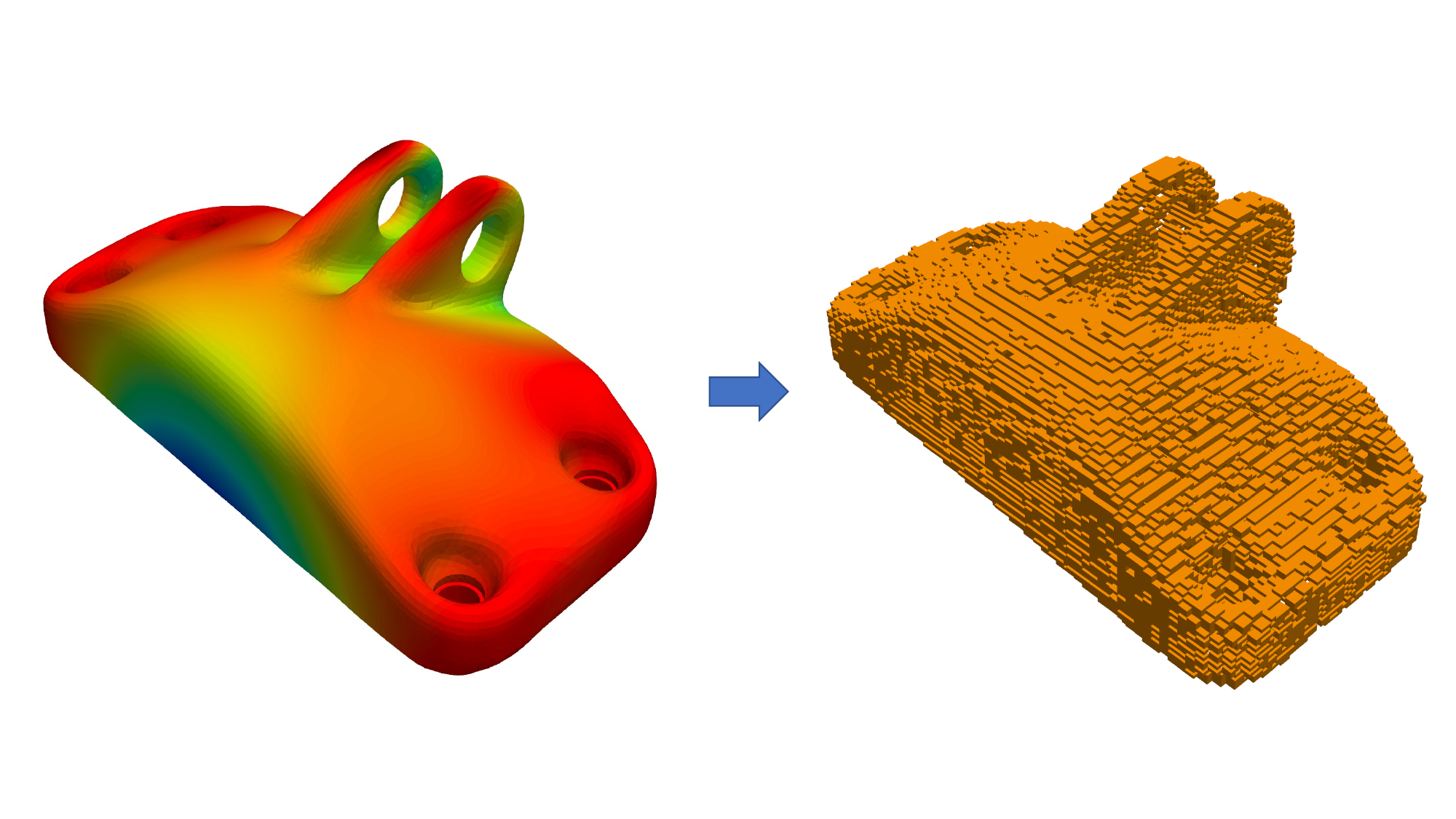}
        \caption{Voxel representation of freeform geometry.}
    \end{subfigure}
    
    \begin{subfigure}[b]{0.45\textwidth}
        \includegraphics[width=\textwidth]{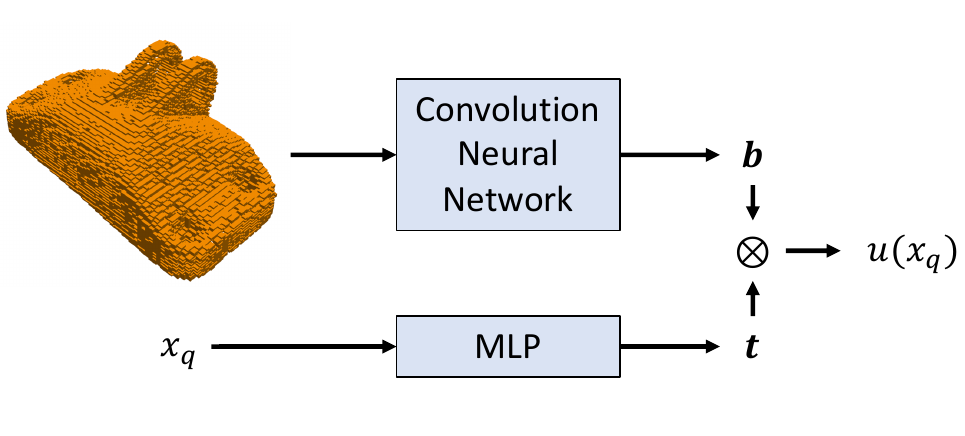}
        \caption{DeepONet with voxel-based branch network.}
    \end{subfigure}
    \begin{subfigure}[b]{0.43\textwidth}
        \includegraphics[width=\textwidth]{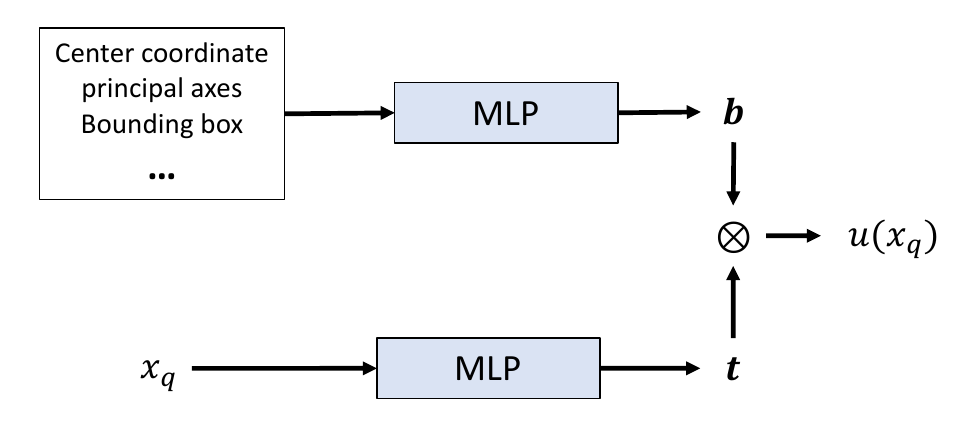}
        \caption{DeepONet with high-level geometric descriptors.}
    \end{subfigure}
    \caption{\footnotesize Branch-trunk model architecture enhancements for freeform geometry.}
    \label{fig.voxel_rep}
\end{figure}

\subsection{Parametric Learning in Geometric Neural Operators}

We investigate two approaches to incorporate parametric design representations into geometric neural operators:

\begin{enumerate}
    \item \textbf{Direct concatenation:} The design parameter vector is directly concatenated with the spatial query point $x_q$ and passed as input to the model.
    
    \item \textbf{Branch-enhanced integration:} A separate MLP branch network processes the parametric vector, while the trunk network encodes the spatial coordinates. The outputs are fused using concatenation, and the resulting features are passed into the geometric operator.
\end{enumerate}

\begin{figure}[h]
    \centering
    \begin{subfigure}[b]{0.37\textwidth}
        \includegraphics[width=\textwidth]{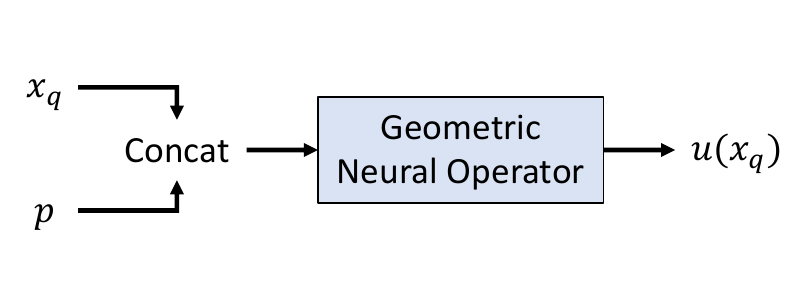}
        \caption{Direct concatenation.}
    \end{subfigure}
    \begin{subfigure}[b]{0.57\textwidth}
        \includegraphics[width=\textwidth]{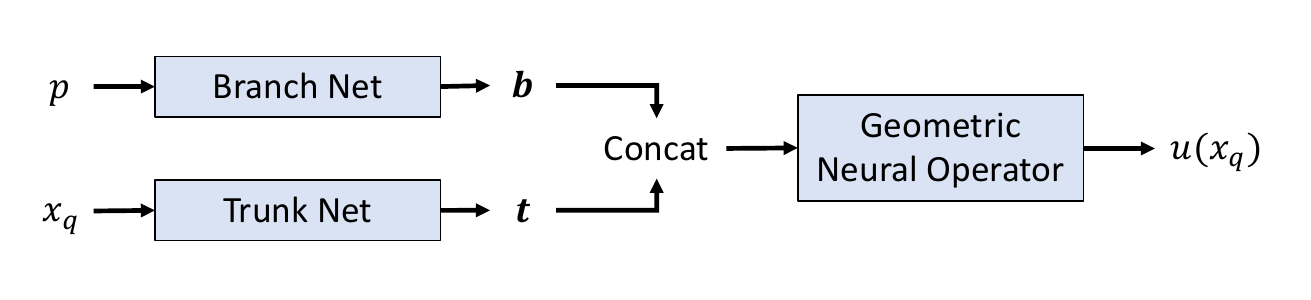}
        \caption{Branch-enhanced geometric neural operator.}
    \end{subfigure}
    \caption{\footnotesize Geometric neural operator architectures incorporating parametric design inputs.}
    \label{fig.geo_param}
\end{figure}

% \subsection{Model Adaptations for Memory-Constrained Datasets}

% For the largest datasets in our study (DrivAerNet and DrivAerNet++), several high-capacity models—such as GINO, GNOT, and Transolver—encounter GPU memory limitations during training. To address this issue, we apply three memory-efficient strategies:

% \begin{itemize}
%     \item \textbf{Farthest point sampling with cross-attention:} A subset of representative control points is selected via farthest point sampling. These control points attend to the full geometry using a cross-attention mechanism, reducing memory usage while preserving global context. During model training and testing, these control points are used for keys and values.
    
%     \item \textbf{Random sampling during training:} To reduce training cost, we sample a subset of the point cloud and only predict the PDE solution on these points during training. Full-resolution inference is performed where all the points are used.
    
%     \item \textbf{Domain partitioning:} The input geometry is partitioned into smaller subregions based on spatial location. Independent models are trained on each partition, enabling large-scale coverage with reduced memory load per model.
% \end{itemize}

% These enhancements enable the deployment of neural operator models at scale while ensuring consistency and fairness across model families.

\section{Numerical experiment \label{Sec.results}}

In this section, we provide a detailed analysis of the benchmarking results across different neural operator families and datasets. Each dataset was split into 60\% for training, 10\% for validation, and 30\% for testing. The models are trained with a learning rate of 0.001, which is adjusted using an adaptive decay schedule to improve convergence stability. Optimization is performed using the Adam optimizer with parameters $\beta_1 = 0.9$ and $\beta_2 = 0.999 $. A batch size of 5 is used, and the training process is carried out for 1000 epochs to ensure sufficient convergence. All experiments are conducted on an NVIDIA A100 GPU with 40GB of memory. The model performance is evaluated using the Mean Absolute Error (MAE) of the predicted fluid velocity and pressure. We use the KDTree~\cite{kdtree} algorithm to construct graph connectivity for point cloud inputs. For grid-based models, the hidden graph resolution is set to a maximum of 100 voxels to ensure sufficient graph density. We follow the original implementation of each model to set its hyperparameters. In each training step, the gradient is computed based on a single graph. If the hidden dimension exceeds available GPU memory, we reduce it accordingly and always use the largest dimension that fits within the memory constraint.

\textcolor{black}{All models are implemented within a unified experimental framework to ensure fair comparisons across datasets. This framework standardizes architectural depth, hidden dimensions, and spatial resolution. Unless otherwise specified, all models use 3 layers. Hidden dimensions are chosen based on model type: 128 for branch-trunk architectures (e.g., DeepONet, Geom-DeepONet), 32 for graph-based operators (e.g., GNO, GNOT, GUNet), 16 for grid-based models (e.g., FNO, FIGConv), and 128 for point-based networks (e.g., GANO, EAGNO). Each model receives input features defined on voxelized or point-sampled geometries, with spatial resolution set by the dataset-specific grid shape. The input parameter dimensions vary across datasets, spanning 0 to 22 for geometric parameters and 0 to 101 for load parameters. For output activation, we apply the sigmoid function to datasets with normalized outputs (Heat sink, Bracket, Bracket-time) and use the identity function for datasets with unbounded or highly variable outputs (JEB, DrivAer, DrivAer++).}

\textcolor{black}{Due to the significant computational cost of training on high-resolution 3D data, it is infeasible to perform exhaustive hyperparameter tuning for each model-dataset pair. To address this, we employ a two-stage configuration strategy. First, we perform a grid search over a subset of each dataset, training and validating on the same split to identify a suitable combination of hidden dimension and layer count. Once the best configuration is selected, it is fixed and applied in all subsequent evaluations for that model. This procedure ensures a balance between computational feasibility and robust, standardized benchmarking across diverse neural operator architectures. The complete spatial configurations of all datasets are summarized in Table~\ref{tab:dataset_grid_config}.}

\begin{table}[h]
\centering
\renewcommand{\arraystretch}{1.0}
\caption{Dataset configuration used across all model evaluations.}
\label{tab:dataset_grid_config}
\begin{tabular}{|c|c|c|c|c|c|c|}
\hline
\textbf{Dataset} & \makecell{\textbf{Geo} \ \textbf{Dim}} & \makecell{\textbf{Load} \ \textbf{Dim}} & \textbf{$x$ range} & \textbf{$y$ range} & \textbf{$z$ range} & \textbf{Grid Shape} \\
\hline
Heat sink & 4 & 0 & [0.0, 1.0] & [0.0, 1.0] & [0.0, 1.0] & (50, 50, 50) \\
Bracket & 3 & 1 & [0.0, 1.0] & [0.0, 1.0] & [0.0, 1.0] & (50, 50, 50) \\
Bracket-time & 0 & 101 & [-0.2, 1.1] & [-0.1, 0.4] & [0.0, 0.2] & (65, 25, 10) \\
JEB & 0 & 0 & [0.0, 1.0] & [0.0, 1.0] & [0.0, 1.0] & (50, 50, 50) \\
DrivAer & 22 & 0 & [-1.2, 4.1] & [-1.1, 1.1] & [0.0, 1.8] & (53, 22, 18) \\
DrivAer++ & 0 & 0 & [-1.2, 4.1] & [-1.1, 1.1] & [0.0, 1.8] & (53, 22, 18) \\
\hline
\end{tabular}
\end{table}

The results are presented in three steps: (1) evaluation of branch–trunk neural operators with parametric inputs, (2) evaluation of general-geometry neural operators (vanilla versions) without parametric information, and (3) evaluation of enhanced variants of both branch–trunk and general-geometry models.

\subsection{Main results}

Table~\ref{tab:branchtrunk_results} compares the performance of branch-trunk neural operators on datasets featuring both explicit parametric geometry representation and free-form geometries. Notably, DCON, with its proposed stacked operator layer, achieves the lowest errors on the Heat sink (0.10\%) and Bracket (1.75\%) datasets. S-NOT, which incorporates temporal attention encoding in the branch network, delivers the best performance on the time-dependent Bracket-time dataset (5.6\%). GANO utilizes parametric representations when available and point cloud representations when parametric inputs are not provided. This explains the significant performance difference observed between the DrivAer and DrivAer++ datasets. We find that for complex vehicle geometries, point cloud representations lead to better performance than parametric representations.

\begin{table}[H]
\centering
\caption{Test errors (\%) of branch-trunk neural operators. Bold indicates best in each column.}
\label{tab:branchtrunk_results}
\renewcommand{\arraystretch}{1.0}
\begin{tabular}{lcccccc}
\toprule
\textbf{Model} & \textbf{Heat sink} & \textbf{Bracket} & \textbf{Bracket-time} & \textbf{JEB} & \textbf{DrivAer} & \textbf{DrivAer++} \\
\midrule
DeepONet        & 0.15\%  & 5.90\%  & 14.8\%  & -     & 53.1\%   & -     \\
Geom-DeepONet   & 0.18\%  & 1.84\%  & 15.1\%  & -     & 51.3\%   & -     \\
S-DeepONet      & -       & -       & 8.6\%  & -     & -       & -     \\
S-NOT      & -       & -       & \textbf{5.6\%}  & -     & -       & -     \\
DCON            & \textbf{0.10\%}  & \textbf{1.75\%}  & 10.7\%  & -     & 47.4\%   & -     \\
GANO            & 0.16\%  & 1.86\%  & 12.9\%  & \textbf{47.1\%}  & \textbf{23.1\%}  & \textbf{21.8\%} \\
\bottomrule
\end{tabular}
\end{table}

Table~\ref{tab:geomodel_noparam} compares the performance of general geometric neural operators on different time-independent datasets \textbf{without parametric inputs}, relying solely on mesh or point cloud representations. A key observation is that no single model consistently outperforms others across all datasets, underscoring the diversity of our benchmark and its utility for evaluating generalization in neural operator design. For datasets with parametric inputs and relatively simple geometries (e.g., Heat sink and Bracket), geometric neural operators are generally less competitive than branch-trunk models. However, for datasets with freeform geometries such as JEB and DrivAer++, geometric models perform significantly better.

In comparing different model types, we observe that GNO consistently yields the highest error across all datasets. Additionally, grid-based models outperform point-based models on the Heat sink and JEB datasets but underperform on Bracket. We hypothesize that this is due to the greater variation in geometry size within the Heat sink and JEB datasets compared to Bracket. These results suggest that grid-based models may be more robust when geometry sizes vary significantly. Conversely, for the DrivAer and DrivAer++ datasets, point-based models consistently achieve superior performance, indicating their strength in capturing complex PDE fields on intricate geometries.

\begin{table}[H]
\centering
\caption{Test errors (\%) of general-geometry models without parametric inputs.}
\label{tab:geomodel_noparam}
\renewcommand{\arraystretch}{1.0}
\begin{tabular}{lccccc}
\toprule
\textbf{Model} & \textbf{Heat sink} & \textbf{Bracket} & \textbf{JEB} & \textbf{DrivAerNet} & \textbf{DrivAer++} \\
\midrule
Best branch-trunk  & \textbf{0.10\% (DCON)} & \textbf{1.75\% (DCON)} & 47.1\% (GANO) & 23.1\% (GANO) & 21.8\% (GANO) \\
GNO                & 5.69\% & 10.14\% & 61.1\% & 45.8\% & 33.8\% \\
EA-GNO             & 6.12\% & 10.88\% & 58.6\% & 48.7\% & 35.8\% \\
MechGraphNet       & 5.94\% & 11.27\% & 62.9\% & 49.6\% & 46.5\% \\
GI-FNO             & 1.21\% & 3.94\%  & 31.2\% & 25.6\% & 19.8\% \\
FigConvUNet        & 0.89\% & 3.80\% & \textbf{29.8\%} & 23.7\% & 18.3\% \\
PointNet           & 6.14\% & 8.75\%  & 40.1\% & 28.5\% & 18.8\% \\
GNOT               & 5.55\% & 3.58\%  & 37.6\% & 17.9\% & 18.3\% \\
Transolver         & 5.23\% & 2.74\%  & 36.6\% & \textbf{16.7\%} & \textbf{17.3\%} \\
\bottomrule
\end{tabular}
\end{table}

\subsection{Performances of enhanced models}

Table~\ref{tab:param_fusion} shows the performance of general-geometry models enhanced through the direct concatenation of parametric inputs. This fusion strategy significantly improves prediction accuracy across most models. For example, \textbf{GINO} improves from 3.94\% to 0.90\% on the Bracket dataset, while \textbf{Transolver} achieves a test error of 0.32\%, outperforming all other models. \textbf{PointNet} and \textbf{GNOT} also benefit substantially from the inclusion of parametric design information. However, for problems involving time-dependent components, incorporating a dedicated temporal module in the machine learning model remains essential.

\begin{table}[H]
\centering
\caption{Test errors (\%) of general-geometry models with parametric input concatenation.}
\label{tab:param_fusion}
\renewcommand{\arraystretch}{1.0}
\begin{tabular}{lccc}
\toprule
\textbf{Model} & \textbf{Heat sink} & \textbf{Bracket} & \textbf{Bracket-time} \\
\midrule
Best branch-trunk   & \textbf{0.10\% (DCON)} & 1.75\% (DCON) & \textbf{5.6\% (S-NOT)} \\
GNO                 & 0.11\% & 1.20\% & 14.2\% \\
EA-GNO              & 0.14\% & 1.79\% & 13.8\% \\
MechGraphNet        & 0.20\% & 23.7\% & 16.5\% \\
GI-FNO              & 0.14\% & 0.90\% & 12.9\% \\
FigConvUNet         & 0.13\% & 0.81\% & 11.8\% \\
PointNet            & 0.12\% & 0.75\% & 10.2\% \\
GNOT                & 0.14\% & 0.45\% & 11.8\% \\
Transolver          & 0.11\% & \textbf{0.32\%} & 10.3\% \\
\bottomrule
\end{tabular}
\end{table}

As shown in Table~\ref{tab:branch_enhanced}, further improvements are achieved by separately processing parametric vectors through a dedicated MLP branch. This branch-enhanced fusion strategy consistently improves performance across all models. \textbf{FigConvUNet} and \textbf{Transolver} exhibit the most significant gains, reducing their Bracket-time errors to 9.8\% and 7.9\%, respectively. The results from Table~\ref{tab:param_fusion} and Table~\ref{tab:branch_enhanced} indicate that, for problems with parametric representations, existing geometric neural operators can achieve accuracy levels comparable to branch-trunk models by effectively fusing local coordinate information with global parametric features. However, their performances still cannot overcome the temporal attention mechanism.

\begin{table}[H]
\centering
\caption{Test errors (\%) with branch-enhanced parametric fusion. Values shown as baseline $\rightarrow$ improved.}
\label{tab:branch_enhanced}
\renewcommand{\arraystretch}{1.0}
\begin{tabular}{lcc}
\toprule
\textbf{Model} & \textbf{Heat sink} & \textbf{Bracket-time} \\
\midrule
Best branch-trunk   & 0.10\% (DCON) & \textbf{5.6\% (S-NOT)} \\
GNO                 & 0.11\% $\rightarrow$ 0.09\% & 14.2\% $\rightarrow$ 11.4\% \\
EA-GNO              & 0.14\% $\rightarrow$ 0.12\% & 13.8\% $\rightarrow$ 11.2\% \\
MechGraphNet        & 0.20\% $\rightarrow$ 0.15\% & 16.5\% $\rightarrow$ 13.0\% \\
GI-FNO              & 0.14\% $\rightarrow$ 0.08\% & 12.9\% $\rightarrow$ 10.2\% \\
FigConvUNet         & \textbf{0.13\% $\rightarrow$ 0.07\%} & 11.8\% $\rightarrow$ 9.8\% \\
PointNet            & 0.12\% $\rightarrow$ 0.10\% & 10.2\% $\rightarrow$ 8.3\% \\
GNOT                & 0.14\% $\rightarrow$ 0.11\% & 11.8\% $\rightarrow$ 8.9\% \\
Transolver          & 0.11\% $\rightarrow$ 0.09\% & 10.3\% $\rightarrow$ 7.9\% \\
\bottomrule
\end{tabular}
\end{table}

Table~\ref{tab:deepo_geom} investigates the effect of different geometry encodings for DeepONet on freeform datasets. High-level feature representations lead to poor accuracy, whereas voxel-based CNN encodings improve performance as voxel resolution increases. However, even with the highest resolution setup (100\textsuperscript{3}), voxel-based DeepONet remains less accurate than the best-performing geometric neural operators. This highlights a fundamental limitation of branch-trunk models when applied to freeform geometries.

\begin{table}[H]
\centering
\caption{Test errors (\%) of DeepONet with different geometric encodings on freeform datasets.}
\label{tab:deepo_geom}
\renewcommand{\arraystretch}{1.0}
\begin{tabular}{lcc}
\toprule
\textbf{Model Variant} & \textbf{JEB} & \textbf{DrivAer++} \\
\midrule
Best model           & \textbf{29.8\% (FigconvUNet)} & \textbf{16.7\% (transolver)} \\
High-level features    & 56.4\%          & 67.2\% \\
Voxel 10$^3$           & 47.4\%          & 46.3\% \\
Voxel 50$^3$           & 42.9\%          & 43.9\% \\
Voxel 100$^3$          & 38.9\%          & 40.7\% \\
\bottomrule
\end{tabular}
\end{table}

We also compare the memory and computational costs of model training by measuring the training time per epoch for each model across all datasets. Table~\ref{tab:training_time_and_params} summarizes these results. Models such as DeepONet, Geom-DeepONet, and PointNet exhibit significantly faster training times, particularly on smaller datasets like Heat sink and Bracket. In contrast, graph-based models are generally slower, reflecting the increased computational complexity associated with graph construction and neighbor searches. Notably, DCON achieves competitive training speeds without compromising predictive accuracy, making it a balanced choice in terms of both efficiency and performance. In terms of the inference, all the model achieved at least 100x faster than Heat sink dataset and at least 1000x faster than other dataset.

\begin{table}[h]
\centering
\caption{Training time per epoch (in seconds) and number of trainable parameters for all models across datasets.}
\label{tab:training_time_and_params}
\begin{tabular}{lcccccc}
\toprule
\textbf{Model} & \textbf{Heat sink} & \textbf{Bracket} & \textbf{Bracket-time} & \textbf{JEB} & \textbf{DrivAer} & \textbf{DrivAer++} \\
\midrule
DeepONet & \makecell{3.68s \\ (133.25K)} & \makecell{20.5s \\ (133.25K)} & \makecell{75.0s \\ (145.66K)} & - & \makecell{2160s \\ (135.55K)} & - \\
Geom-DeepONet & \makecell{3.78s \\ (100.22K)} & \makecell{16.6s \\ (100.35K)} & \makecell{79.4s \\ (125.57K)} & - & \makecell{900s \\ (102.53K)} & - \\
S-DeepONet & - & - & \makecell{74.3s \\ (427.24K)} & - & - & - \\
DCON & \makecell{3.94s \\ (100.35K)} & \makecell{21.0s \\ (100.35K)} & \makecell{73.4s \\ (112.77K)} & - & \makecell{2320s \\ (102.66K)} & - \\
GANO & \makecell{3.79s \\ (215.81K)} & \makecell{19.6s \\ (282.11K)} & \makecell{78.0s \\ (320.64K)} & \makecell{389s \\ (215.68K)} & \makecell{2057s \\ (218.11K)} & \makecell{750s \\ (215.68K)} \\
\midrule
GNO & \makecell{45.0s \\ (212.16K)} & \makecell{230s \\ (212.19K)} & \makecell{372s \\ (215.39K)} & \makecell{1200s \\ (212.16K)} & \makecell{3600s \\ (212.16K)} & \makecell{3600s \\ (212.16K)} \\
EA-GNO & \makecell{52.0s \\ (212.16K)} & \makecell{408s \\ (212.19K)} & \makecell{375s \\ (215.39K)} & \makecell{419s \\ (212.16K)} & \makecell{3800s \\ (212.16K)} & \makecell{3800s \\ (212.16K)} \\
\midrule
GINO & \makecell{48.0s \\ (162.15M)} & \makecell{211s \\ (162.15M)} & \makecell{98.4s \\ (26.10M)} & \makecell{194s \\ (76.11M)} & \makecell{4670s \\ (38.16M)} & \makecell{467s \\ (38.16M)} \\
FigconvUNet & \makecell{42.0s \\ (79.43M)} & \makecell{197s \\ (79.43M)} & \makecell{89.4s \\ (77.75M)} & \makecell{235s \\ (99.86M)} & \makecell{4320s \\ (81.73M)} & \makecell{4120s \\ (81.73M)} \\
\midrule
PointNet & \makecell{3.70s \\ (247.94K)} & \makecell{16.9s \\ (248.06K)} & \makecell{83.9s \\ (260.87K)} & \makecell{17.4s \\ (247.94K)} & \makecell{771s \\ (247.94K)} & \makecell{750s \\ (247.94K)} \\
GNOT & \makecell{43.0s \\ (3.67M)} & \makecell{235s \\ (3.67M)} & \makecell{72.9s \\ (3.88M)} & \makecell{283s \\ (3.67M)} & \makecell{3600s \\ (3.88M)} & \makecell{3420s \\ (3.88M)} \\
Transolver & \makecell{49.0s \\ (585.30K)} & \makecell{228s \\ (585.56K)} & \makecell{75.3s \\ (611.16K)} & \makecell{297s \\ (585.30K)} & \makecell{3600s \\ (581.10K)} & \makecell{3600s \\ (581.10K)} \\
\bottomrule
\end{tabular}
\end{table}

Combining all the analyses above, we summarize our key findings as follows:

\begin{itemize}
    \item \textbf{For problems with parametric representations}, branch-trunk models should be the first choice due to their low training cost and strong performance on structured geometries. For example, DCON consistently achieves the best results on the Heat sink (0.10\%) and Bracket (1.75\%) datasets (Table~\ref{tab:branchtrunk_results}), while maintaining fast per-epoch training times (Table~\ref{tab:training_time_and_params}).
    \item \textbf{Even when parametric inputs are available}, point-based models may be preferable when dealing with complex geometries or solutions with abrupt spatial changes. As shown in Table~\ref{tab:branchtrunk_results}, transolver outperforms all other models on the DrivAer (16.7\%) and DrivAer++ (17.3\%) datasets, both of which feature freeform vehicle geometries.
    \item \textbf{When geometry size varies significantly across the dataset}, grid-based neural operators provide more robust performance. For instance, FigConvUNet achieves the best results on the Heat sink (0.89\%) and JEB (29.8\%) datasets (Table~\ref{tab:geomodel_noparam}), where geometry scale variation is pronounced.
    \item \textbf{For time-dependent problems}, \textcolor{black}{Models equipped with explicit temporal encoding mechanisms can achieve strong performance, even when the overall architecture is as simple as a branch-trunk design. This is evident in the superior performance of S-DeepONet (8.6\% error) compared to all other models on the Bracket-time dataset (Table~\ref{tab:param_fusion}), owing to its GRU-based temporal encoder. The performance of SNOT further validate our conclusion.}
\end{itemize}

\section{Conclusion \label{Sec.conclusion}}

In this study, we introduced a comprehensive benchmarking framework for evaluating neural operator architectures on six industry-scale 3D engineering design datasets, covering diverse physical domains such as thermal analysis, elasto-plasticity, time-dependent mechanics, and computational fluid dynamics. Our benchmark encompassed twelve neural operator models spanning four architectural categories: branch-trunk, graph-based, grid-based, and point-based. To ensure a fair comparison across domains and geometries, we implemented enhancements that extend branch-trunk models to freeform geometries and enable geometric neural operators to incorporate parametric inputs.

Our findings demonstrate that no single neural operator architecture consistently outperforms others across all tasks. Instead, model performance is closely tied to the alignment between architectural design, input representation, and the underlying physics of each problem. Branch-trunk models—particularly DCON and S-DeepONet—excel in problems with structured parametric inputs and temporal dynamics. Grid-based models like FigConvNet show strong performance on geometrically complex domains due to their spatial regularity and scalability. Point-based models such as Transolver perform robustly across both structured and unstructured settings, leveraging their physics-aware attention mechanisms for flexible representation learning. Despite the breadth of our study, several limitations remain:

\begin{itemize}
    \item \textbf{Model Coverage:} While we benchmarked twelve models, the landscape of neural operator architectures is evolving rapidly. Future work should explore recently proposed models, particularly those designed for multi-physics coupling, uncertainty-aware learning, and advanced spatiotemporal modeling.
    \item \textbf{Dataset Scope:} Our benchmark includes six datasets covering key engineering domains, but it remains limited in breadth. Extending the framework to include problems such as fluid-structure interaction (FSI), contact mechanics, turbulence modeling, and additive manufacturing would enhance generalizability.
    \item \textbf{Scalability and Deployment:} Although we proposed deployment-oriented modifications, our current evaluation does not fully address scalability in real-world settings. Future studies should assess performance under constraints such as limited GPU memory, multi-node distributed environments, and hardware heterogeneity (e.g., CPU vs GPU inference). Testing on high-performance machines with larger memory would be particularly useful for evaluating deployment viability.
    \item \textbf{Interpretability and Robustness:} The interpretability of neural operator predictions and their robustness to noise, mesh perturbations, or out-of-distribution (OOD) inputs remain open challenges. These aspects are especially important for trustworthy adoption in safety-critical engineering applications.
    \item \textbf{Uncertainty Quantification:} Our current benchmarking framework does not account for uncertainty in model predictions. Incorporating principled uncertainty quantification methods—such as Bayesian neural operators or ensemble-based approaches~\cite{pivae}—is essential for risk-aware design and decision-making.
    \item \textbf{Data Efficiency and Generalization:} While our benchmark emphasizes accuracy and scalability, further analysis is needed to evaluate data efficiency and model generalization under low-data regimes or with varying boundary conditions. This is particularly relevant for applications with expensive or sparse simulation data.
\end{itemize}

\section*{Acknowledgment}
This work used the \textit{Delta} advanced computing resources provided by the University of Illinois Urbana-Champaign (UIUC) through the National Center for Supercomputing Applications (NCSA). 
We gratefully acknowledge the support of these resources.

\newpage

\bibliography{references}
\bibliographystyle{plainnat}

\end{document}